\documentclass[12pt]{article}

\usepackage{lipsum} % Package to generate dummy text throughout this template

\usepackage[T1]{fontenc} % Use 8-bit encoding that has 256 glyphs
\usepackage{microtype} % Slightly tweak font spacing for aesthetics

\usepackage[margin=1in]{geometry}

\usepackage{multicol} % Used for the two-column layout of the document
\usepackage[hang, small,labelfont=bf,up,textfont=it,up]{caption} % Custom captions under/above floats in tables or figures
\usepackage{booktabs} % Horizontal rules in tables
\usepackage{multirow}
\usepackage{siunitx}
\usepackage{float} % Required for tables and figures in the multi-column environment - they need to be placed in specific locations with the [H] (e.g. \begin{table}[H])
\usepackage{hyperref} % For hyperlinks in the PDF

\usepackage{lettrine} % The lettrine is the first enlarged letter at the beginning of the text
\usepackage{paralist} % Used for the compactitem environment which makes bullet points with less space between them
\usepackage{cite} % for hyphenated lists of citations
\usepackage{apacite} % for APA-style citations
\usepackage{natbib}

\usepackage{amsmath,amsfonts,amsthm, textcomp} % Math packages
\usepackage{bm}

\makeatletter
\renewcommand*\env@matrix[1][\arraystretch]{%
  \edef\arraystretch{#1}%
  \hskip -\arraycolsep
  \let\@ifnextchar\new@ifnextchar
  \array{*\c@MaxMatrixCols c}}
\makeatother

\usepackage{graphicx}
\usepackage{float}
\usepackage{booktabs} % Horizontal rules in tables
 \usepackage{ amssymb }
 
 \usepackage{todonotes}
\usepackage{abstract} % Allows abstract customization
\usepackage{titlesec} % Allows customization of titles
\titleformat{\section}[block]{\large\scshape\centering}{\thesection.}{1em}{} % Change the look of the section titles
\titleformat{\subsection}[block]{\large}{\thesubsection.}{1em}{} % Change the look of the section titles

\usepackage{fancyhdr} % Headers and footers
\usepackage{graphicx}

\usepackage{xcolor}

 \numberwithin{equation}{section}

\title{\vspace{-15mm}\fontsize{18pt}{19pt}\selectfont\textbf{Sensitivity Analysis for Unmeasured Confounding in Meta-Analyses}} % Article title

\author
{Maya B. Mathur$^{1, 2\ast}$ and Tyler J. VanderWeele$^{1,3}$\\
\\
\small{$^{1}$ Department of Biostatistics, Harvard T. H. Chan School of Public Health, Boston, MA, USA}\\
\small{$^{2}$Quantitative Sciences Unit, Stanford University, Palo Alto, CA, USA}\\
\small{$^{3}$Department of Epidemiology, Harvard T. H. Chan School of Public Health, Boston, MA, USA}\\
\\
}

\begin{document}

\maketitle % Insert title

\noindent $^\ast$Corresponding author:
\\mmathur@stanford.edu
\\Quantitative Sciences Unit (c/o Inna Sayfer)
\\1070 Arastradero Road
\\Palo Alto, CA
\\94305

\thispagestyle{fancy} % All pages have headers and footers

%----------------------------------------------------------------------------------------
%	TABLE OF CONTENTS & LISTS OF FIGURES AND TABLES
%----------------------------------------------------------------------------------------

\setcounter{tocdepth}{4} % Set the depth of the table of contents to show sections and subsections only

\newpage

%\tableofcontents % Print the table of contents

\newpage

%----------------------------------------------------------------------------------------
%	ABSTRACT
%----------------------------------------------------------------------------------------

\renewcommand{\abstractname}{Summary} % changes title of abstract
\begin{abstract}
\noindent {
Random-effects meta-analyses of observational studies can produce biased estimates if the synthesized studies are subject to unmeasured confounding. We propose sensitivity analyses quantifying the extent to which unmeasured confounding of specified magnitude could reduce to below a certain threshold the proportion of true effect sizes that are scientifically meaningful. We also develop converse methods to estimate the strength of confounding capable of reducing the proportion of scientifically meaningful true effects to below a chosen threshold. These methods apply when a ``bias factor'' is assumed to be normally distributed across studies or is assessed across a range of fixed values. Our estimators are derived using recently proposed sharp bounds on confounding bias within a single study that do not make assumptions regarding the unmeasured confounders themselves or the functional form of their relationships to the exposure and outcome of interest. We provide an \texttt{R} package, \texttt{ConfoundedMeta}, and a freely available online graphical user interface that compute point estimates and inference and produce plots for conducting such sensitivity analyses. These methods facilitate principled use of random-effects meta-analyses of observational studies to assess the strength of causal evidence for a hypothesis. 
} 
\\*

\end{abstract}

\noindent\textbf{Key words:} Bias; Confounding; Meta-analysis; Observational studies; Sensitivity analysis
%\pagebreak
%\section{Notes to self}
%\begin{itemize}
% \end{itemize}

%----------------------------------------------------------------------------------------
%	Introduction
%----------------------------------------------------------------------------------------
\pagebreak
\section{Introduction}
Meta-analyses can be indispensable for assessing the overall strength of evidence for a hypothesis and for precisely estimating effect sizes through aggregation of estimates. However, conclusions drawn from meta-analyses are only as reliable as the synthesized studies themselves; systematic bias in the meta-analyzed studies typically produces bias in the pooled point estimate \citep{egger}. A common source of bias is unmeasured confounding \citep{shrier}, which is our focus in this paper. When eliminating such bias by restricting attention to well-designed randomized studies is infeasible because the exposure cannot be randomized, an attractive option is to conduct sensitivity analyses assessing the extent to which unmeasured confounding of varying magnitudes could have compromised the results of the meta-analysis.

Existing sensitivity analyses for confounding bias or other internal biases in meta-analysis estimate a bias-corrected pooled point estimate by directly incorporating one or more bias parameters in the likelihood and placing a Bayesian prior on the distribution of these parameters \citep{welton,mccandless}. An alternative frequentist approach models bias as additive or multiplicative within each study and then uses subjective assessment to elicit study-specific bias parameters \citep{turner}. Although useful, these approaches typically require strong assumptions on the nature of unmeasured confounding (for example, requiring a single binary confounder), rely on the arbitrary specification of additive or multiplicative effects of bias, or require study-level estimates rather than only meta-analytic pooled estimates. Furthermore, the specified bias parameters do not necessarily lead to precise practical interpretations.

An alternative approach is to analytically bound the effect of unmeasured confounding on the results of a meta-analysis. To this end, bounding methods are currently available for point estimates of individual studies. We focus on sharp bounds derived by \citet{ding} because of their generality and freedom from assumptions regarding the nature of the unmeasured confounders or the functional forms of their relationships with the exposure of interest and outcome. This approach subsumes several earlier approaches \citep{cornfield,schlesselman,flanders} and relies on only two simple sensitivity parameters representing the strength of association of the unmeasured confounders with, firstly, the exposure and, secondly, the outcome.

The present paper extends these analytic bounds for single studies to the meta-analytic setting. Using standard estimates from a random-effects meta-analysis and intuitively interpretable sensitivity parameters on the magnitude of confounding, these results enable inference about the size of the true, unconfounded effects in a potentially heterogeneous population of studies. That is, we can select a minimum threshold of scientific importance for the magnitude of the true effect in any given study. If sensitivity analysis for unmeasured confounding indicates that too few studies in the meta-analysis have a true effect stronger than this threshold, then arguably the results of the meta-analysis are not robust to confounding, and scientifically meaningful causal conclusions are not warranted despite the observed point estimate. To this end, we develop estimators that answer the questions: ``What proportion of studies would have a true effect size stronger than $q$ in the presence of unmeasured confounding of a specified strength?'' and ``How severe would unmeasured confounding need to be to reduce to less than $r$ the proportion of studies with true effect size stronger than $q$?''. This approach to sensitivity analysis is essentially a meta-analytic extension of a recently proposed metric (the ``E-value'') that quantifies, for a single study, the minimum confounding bias capable of reducing the true effect to a chosen threshold \citep{evalue}. We provide and demonstrate use of an \texttt{R} package (\texttt{ConfoundedMeta}) and a free, interactive online user interface for conducting such analyses and creating plots.

%----------------------------------------------------------------------------------------
%	Setting and Notation
%----------------------------------------------------------------------------------------

%\section{Setting and Notation}
%\label{notation}

% internal citations:
% \citet*{ding}

%----------------------------------------------------------------------------------------
%	Prior Work
%----------------------------------------------------------------------------------------

\section{Existing Bounds on Confounding Bias in a Single Study}

% measure theory section of Ding: pg 31
% general nonneg: pg 61

\citet{ding} developed bounds for a single study as follows. Let $X$ denote a binary exposure, $Y$ a binary outcome, $Z$ a vector of measured confounders, and $U$ one or more unmeasured confounders. Let:
$$RR_{XY | z}^c = \frac{ P \left( Y = 1 \; \vert \; X=1, Z=z \right) }{ P \left( Y = 1 \; \vert \; X=0, Z=z \right) }$$
be the confounded relative risk ($RR$) of $Y$ for $X=1$ versus $X=0$ conditional or stratified on the measured confounders $Z=z$.

Let its true, unconfounded counterpart standardized to the population be:
$$RR_{XY | z}^t = \frac{ \sum_{u} P \left( Y =1 \; \vert \; X=1, Z=z, U=u \right) P \left( U = u \; \vert \; Z=z \right) }{ \sum_{u} P \left( Y \; \vert \; X=0, Z=z, U=u \right) P \left( U = u \; \vert \; Z=z \right) }$$

(Throughout, we use the term ``true'' as a synonym for ``unconfounded'' or ``causal'' when referring to both sample and population quantities. Also, henceforth, we condition implicitly on $Z=z$, dropping the explicit notation for brevity.)

Let $RR_{Xu} = P \left( U = u \; \vert \; X = 1 \right) / P \left( U = u \; \vert \; X = 0 \right)$. Define the first sensitivity parameter as $RR_{XU} = \text{max}_u \left( RR_{Xu} \right)$; that is, the maximal relative risk of $U=u$ for $X=1$ versus $X=0$ across strata of $U$. (If $U$ is binary, this is just the relative risk relating $X$ and $U$.) Next, for each stratum $x$ of $X$, define a relative risk of $Y$ on $U$, maximized across all possible contrasts of $U$:
$$RR_{UY \vert X=x } = \frac{ \max_{u} P \left( Y =1 \vert X=x, U=u \right) }{ \min_{u} P \left( Y = 1 \vert X=x, U=u \right) }, x \in \{ 0, 1 \}$$
Define the second sensitivity parameter as $RR_{UY} = \max \left( RR_{UY \vert X=0}, RR_{UY \vert X=1}\right)$. That is, considering both strata of $X$, it is the largest of the maximal relative risks of $Y$ on $U$ conditional on $X$. Then, \citet{ding} showed that a sharp bound for the true effect is: 
\begin{align}
RR_{XY}^t &\ge RR_{XY}^c / \frac{ RR_{XU} \cdot RR_{UY} }{ RR_{XU} + RR_{UY} - 1}
\label{boundfactor}
\end{align}
where we will refer to the ``bias factor'' $\frac{ RR_{XU} \cdot RR_{UY} }{ RR_{XU} + RR_{UY} - 1}$ as $B$.

If the two sensitivity parameters are equal ($RR_{XU} = RR_{UY}$), then to produce a bias factor $B$, each must exceed $B + \sqrt{B^2 - B}$ \citep{ding}. Thus, a useful transformation of $B$ is the ``confounding strength scale'', $g$, which is the minimum size of $RR_{XU}$ and $RR_{UY}$ under the assumption that they are equal:
\begin{align}
g &= B + \sqrt{B^2 - B} \; \; \; \Leftrightarrow \; \; \; B = \frac{g^2}{2g - 1}
\label{confstrength}
\end{align}

If $RR_{XY}^c < 1$ (henceforth the ``apparently preventive case''), then Equation (\ref{boundfactor}) becomes \citep{ding}:
\begin{align*}
RR_{XY}^t &\le RR_{XY}^c \cdot \frac{ RR_{XU}^{*} \cdot RR_{UY} }{ RR_{XU}^{*}  + RR_{UY} - 1}
\end{align*}
where $RR_{XU}^{*} = \max_u \left( RR_{Xu}^{-1} \right)$, i.e., the maximum of the inverse relative risks, rather than the relative risks themselves. Thus, $B$ remains $\ge 1$, and we have $RR_{XY}^t \ge RR_{XY}^c$.  

Although these results hold for multiple confounders, in the development to follow, we will use a single, categorical unmeasured confounder for clarity. However, all results can easily be interpreted without assumptions on the type of exposure and unmeasured confounders, for instance by interpreting the relative risks defined above as ``mean ratios'' \citep{ding}.

%----------------------------------------------------------------------------------------
%	Setting
%----------------------------------------------------------------------------------------

\section{Random-Effects Meta-Analysis Setting}

In this paper, we use the aforementioned analytic bounds to derive counterparts for the random-effects meta-analysis model with the standard Dersimonian-Laird point estimate. This model assumes that each of $k$ studies measures a potentially unique effect size $M$, such that $M \sim_{iid} N(\mu, V)$ for a grand mean $\mu$ and variance $V$. Let $y_i$ be the point estimate of the $i^{th}$ study and $\sigma^2_{i}$ the within-study variance (with the latter assumed fixed and known).

Analysis proceeds by first estimating $V$ via one of many possible estimators, denoted $\tau^{2}$ \citep{veroniki}, then estimating $\mu$ via a weighted mean defined as:
\begin{align*}
\widehat{y}_{R} &= \frac{\sum_{i=1}^k w_{i} \; y_{i} }{ \sum_{i=1}^k w_{i} }
\end{align*}

The weights are inversely proportional to the total variance of each study (a sum of the between-study variance and the within-study variance), such that $w_{i} = 1 / \left( \tau^{2} + \sigma^{2}_{i} \right)$.

%----------------------------------------------------------------------------------------
%	Main Result
%----------------------------------------------------------------------------------------

\section{Main Results}
\label{main_results_section}

Consider $k$ studies measuring relative risks with confounded population effect sizes on the log-$RR$ scale, denoted $M^c$, such that $M^c \sim N(\mu^c, V^c)$. (Other outcome measures are considered briefly in the Discussion.) Let the corresponding true effects be $M^t$ with expectation $\mu^t$ and variance $V^t$. Let $\widehat{y}_{R}^c$ be the standard inverse-variance-weighted random effects point estimate and $\tau^2_c$ be a heterogeneity estimate, both computed from the confounded data. Consider the bias factor on the log scale, $B^{*} = \log \left( \frac{ RR_{XU} \cdot RR_{UY} }{ RR_{XU} + RR_{UY} - 1} \right)$, and allow it to vary across studies under the assumption that $B^{*} \sim N\left( \mu_{B^{*}}, \sigma^{2}_{B^{*}} \right)$ independently of $M^t$. That is, we assume that the bias factor is independent of the true effects but not the confounded effects: naturally, studies with larger bias factors will tend to obtain larger effect sizes. The normality assumption on the bias factor holds approximately if, for example, its components ($RR_{XU}$ and $RR_{UY}$) are identically and independently normal with relatively small variance (Web Appendix). We now develop three estimators enabling sensitivity analyses.

%-----------------------------
\subsection{Proportion of studies with large effect sizes as a function of the bias factor}
\label{proplarge}

For an \textbf{apparently causative relative risk} ($\widehat{y}_{R}^c > 0$, or equivalently the confounded pooled $RR$ is greater than $1$), define $p(q) = P \left( M^t > q \right)$ for any threshold $q$, i.e., the proportion of studies with true effect sizes larger than $q$. Then a consistent estimator of $p(q)$ is:
\begin{align*}
\widehat{p}(q) &= 1 - \Phi \left( \frac{ q + \mu_{B^{*}} - \widehat{y}^{c}_{R} }{ \sqrt{ \tau^{2}_c - \sigma^{2}_{B^{*}} } } \right), \; \tau^{2}_c > \sigma^{2}_{B^{*}}
\end{align*}
where $\Phi$ denotes the standard normal cumulative distribution function. In the special case in which the bias factor is fixed to $\mu_{B^{*}}$ across all studies, the same formula applies with $\sigma^{2}_{B^{*}} = 0$. 

Many common choices of heterogeneity estimators, $\tau^2_c$, are asymptotically independent of $\widehat{y}_R^c$ (Web Appendix), an assumption used for all standard errors in the main text. Results relaxing this assumption appear throughout the Web Appendix. An application of the delta method thus yields an approximate standard error:
\begin{align*}
\widehat{\text{SE}}\left( \widehat{p}(q) \right) &\approx \sqrt{ \frac{ \widehat{ \text{Var}}\left( \widehat{y}^{c}_{R} \right) }{  \tau^{2}_c - \sigma^{2}_{B^{*}} } + \frac{ \widehat{\text{Var}}\left( \tau^{2}_c \right) \left( q + \mu_{B^{*} } - \widehat{y}_{R}^{c} \right)^{2} }{ 4 \left( \tau^{2}_c - \sigma^{2}_{B^{*}} \right)^{3} } } \cdot \phi \left( \frac{ q + \mu_{B^{*} } - \widehat{y}_{R}^{c} }{ \sqrt{ \tau^{2}_c - \sigma^{2}_{B^{*}} } } \right)
\end{align*}
where $\phi$ denotes the standard normal density function. (If $\tau^{2}_c \le \sigma^{2}_{B^{*}}$, leaving one of the denominators undefined, this indicates that there is so little observed heterogeneity in the confounded effect sizes that, given the specified bias distribution, $V^t$ is estimated to be less than $0$. Therefore, attention should be limited to a range of values of $\sigma^{2}_{B^{*}}$ such that $\tau^{2}_c > \sigma^{2}_{B^{*}}$.)

For an \textbf{apparently preventive relative risk} ($\widehat{y}_{R}^c < 0$ or the confounded pooled $RR$ is less than $1$), define instead $p(q) = P \left( M^t < q \right)$, i.e., the proportion of studies with true effect sizes less than $q$. Then a consistent estimator is:
\begin{align*}
\widehat{p}(q) &= \Phi \left( \frac{ q - \mu_{B^{*}} - \widehat{y}^{c}_{R} }{ \sqrt{ \tau^{2}_c - \sigma^{2}_{B^{*}} } } \right), \; \tau^{2}_c > \sigma^{2}_{B^{*}}
\end{align*}

with approximate standard error:
\begin{align}
\widehat{\text{SE}}\left( \widehat{p}(q) \right) &=\sqrt{ \frac{ \widehat{\text{Var}}\left( \widehat{y}^{c}_{R} \right) }{  \tau^{2}_c - \sigma^{2}_{B^{*}} } + \frac{ \widehat{\text{Var}}\left( \tau^{2}_c  \right) \left( q - \mu_{B^{*} } - \widehat{y}_{R}^{c} \right)^{2} }{ 4 \left( \tau^{2}_c - \sigma^{2}_{B^{*}} \right)^{3} } } \cdot \phi \left( \frac{ q - \mu_{B^{*} } - \widehat{y}_{R}^{c} }{ \sqrt{ \tau^{2}_c - \sigma^{2}_{B^{*}} } } \right)
\end{align}

Because $\widehat{p}(q)$ is monotonic in $\sigma^2_{B^{*}}$, the homogeneous bias case (i.e., $\sigma^2_{B^{*}}=0$) provides either an upper or lower bound on $\widehat{p}(q)$ (Table \ref{homo-bias}). We later return to the practical utility of these results.

%------------------------------------------
\subsection{Bias factor required to reduce proportion of large effect sizes to a threshold}

Conversely, we might consider the minimum common bias factor (on the $RR$ scale) capable of reducing to less than $r$ the proportion of studies with true effect exceeding $q$. We accordingly define $T(r,q) = B : P \left( M^t > q \right) = r$ to be this quantity, with $B$ taken to be constant across studies. (Note that taking $B$ to be constant does not necessarily imply that the unmeasured confounders themselves are identical across studies.) Then for an \textbf{apparently causative relative risk}, a consistent estimator for the the minimum common bias capable of reducing to less than $r$ the proportion of studies with effects surpassing $q$ is:
\begin{align}
\widehat{T}(r,q) &= \exp \Big\{ \Phi^{-1}(1-r) \sqrt{ \tau^{2}_c } - q + \widehat{y}^{c}_{R} \Big\}
\label{T_causative}
\end{align}
with approximate standard error:
\begin{align}
\widehat{\text{SE}}\left( \widehat{T}(r,q) \right) &= \exp \Bigg\{ \sqrt{ \tau^2_c } \left( \Phi^{-1} (1-r) \right) - q + \widehat{y}^{c}_{R} \Bigg\} \sqrt{ \widehat{\text{Var}}\left( \widehat{y}^{c}_{R} \right) + \frac{ \widehat{\text{Var}}\left( \tau^2_c \right) \left( \Phi^{-1} (1-r) \right)^2 }{ 4 \tau^2_c } }
\end{align}

For an \textbf{apparently preventive relative risk}, we can instead consider the minimum common bias factor (on the $RR$ scale) capable of reducing to less than $r$ the proportion of studies with true effect less than $q$, thus defining $T(r,q) = B : P \left( M^t < q \right) = r$. Then a consistent estimator is:
\begin{align}
\widehat{T}(r,q) &=  \exp \Big\{ q - \widehat{y}^{c}_{R} - \Phi^{-1}(r) \sqrt{ \tau^{2}_c } \Big\}
\end{align}

with approximate standard error:
\begin{align}
\widehat{\text{SE}}\left( \widehat{T}(r,q) \right) &= \exp \Bigg\{ q - \widehat{y}^{c}_{R} - \sqrt{ \tau^2_c } \left( \Phi^{-1} (r) \right) \Bigg\} \sqrt{ \widehat{\text{Var}}\left( \widehat{y}^{c}_{R} \right) + \frac{ \widehat{\text{Var}}\left( \tau^2_c \right) \left( \Phi^{-1} (r) \right)^2 }{ 4 \tau^2_c } }
\end{align}

%------------------------------------------
\subsection{Confounding strength required to reduce proportion of large effect sizes to a threshold}

Under the assumption that the two components of the common bias factor are equal as in Equation \ref{confstrength}, such that $g = RR_{XU} = RR_{UY}$, the bias can alternatively be parameterized on the confounding strength scale. Consider the minimum confounding strength required to lower to less than $r$ the proportion of studies with true effect exceeding $q$ and accordingly define $G(r,q) = g : P \left( M^t > q \right) = r$. For both the \textbf{apparently causative and the apparently preventive cases}, an application of Equation \ref{confstrength} yields:
\begin{align}
\widehat{G}(r,q) &= \widehat{T}(r,q) + \sqrt{ \left( \widehat{T}(r,q) \right)^2 - \widehat{T}(r,q) }
\end{align}

with approximate standard error:
\begin{align*}
\widehat{ \text{SE} } \left( \widehat{G}(r,q) \right) &= \widehat{ \text{SE} } \left( \widehat{T}(r,q) \right) \cdot \left( 1 + \frac{2\widehat{T}(r,q) - 1}{ 2 \sqrt{\widehat{T}(r,q)^2 - \widehat{T}(r,q)} } \right)
\end{align*}

%----------------------------------------------------------------------------------------
\section{Practical Use and Interpretation}

The estimators $\widehat{p}(q), \widehat{T}(r,q)$, and $\widehat{G}(r,q)$ enable several types of sensitivity analysis. Firstly, $\widehat{p}(q)$ can be computed over a range of values of $\mu_{B^{*}}$ and $\sigma^2_{B^{*}}$. If $\widehat{p}(q)$ remains large for even large values of $\mu_{B^{*}}$, this indicates that even if the influence of unmeasured confounding were substantial, a large proportion of studies nevertheless would have true effects of scientifically meaningful magnitudes. Similarly, $\widehat{T}(r,q)$ and $\widehat{G}(r,q)$ can be computed for $r$ representing a ``large enough'' proportion of studies to warrant scientific interest; large values would again lead to the conclusion that results of the meta-analysis are relatively robust to unmeasured confounding. For example, by choosing $q=\log (1.10)$ and $r=0.20$ and computing $\widehat{T}(r,q) = 2.50$ (equivalently, $\widehat{G}(r,q) = 4.44$), one might conclude: ``The results of this meta-analysis are relatively robust to unmeasured confounding, insofar as a bias factor of $2.50$ on the relative risk scale (e.g., a confounder associated with the exposure and outcome by risk ratios of $4.44$ each) in each study would be capable of reducing to less than $20\%$ the proportion of studies with true relative risks greater than $1.10$, but weaker confounding could not do so.'' On the other hand, small values of $\widehat{p}(q), \widehat{T}(r,q)$, and $\widehat{G}(r,q)$ indicate that only weak unmeasured confounding would be required to reduce the effects to a scientifically unimportant level; the meta-analysis would therefore not warrant strong scientific conclusions regarding causation.

A general guideline might be to use $q = \log 1.10$ for an apparently causative relative risk or $q = \log 0.90$ for an apparently preventive relative risk. When the number of studies, $k$, is large (for example, $\ge 10$), one might require at least 10\% of studies ($r=0.10$) to have effect sizes above $q$ for results to be of scientific interest. For $k<10$, one might select a higher threshold, such as $r=0.20$ (thus requiring at least 20\% of studies to have effects more extreme than, for example, $\log 1.10$). Of course, these guidelines can and should be adapted based on the substantive application. Furthermore, note that the amount of bias that would be considered ``implausible'' must be determined with attention to the design quality of the synthesized studies: a large bias factor may be plausible for a set of studies with poor confounding control and with high potential for unmeasured confounding, but not for a set of better-designed studies in which the measured covariates already provide good control of confounding.

Sensitivity analyses based on $\widehat{p}(q)$ should be reported for a wide range of values for $\mu_{B^{*}}$ and with $\sigma^2_{B^{*}}$ ranging from $0$ to somewhat less than $\tau^2_c$. The bounds achieved when $\sigma^2_{B^{*}} = 0$ (Table \ref{homo-bias}) can provide useful conservative analyses. For example, for $\widehat{y}_R^c > 0$ and $q > \widehat{\mu}^t$, the $\sigma^2_{B^{*}}=0$ case provides an upper bound on $\widehat{p}(q)$. When concluding that results are not robust to unmeasured confounding, the analysis with $\sigma^2_{B^{*}}=0$ is therefore conservative in that fewer true effect sizes would surpass $q$ under heterogeneous bias. For example, if we calculated $\widehat{T} \left(r=0.20,q=\log 1.10 \right) = 1.20$, then an analysis like this would yield conclusions such as: ``The results of this meta-analysis are relatively sensitive to unmeasured confounding. Even a bias factor as small as $1.20$ in each study would reduce to less than 20\% the proportion of studies with true relative risks greater than $1.10$, and if the bias in fact varied across studies, then even fewer studies would surpass this effect size threshold.''

%----------------------------------------------------------------------------------------
%	Applied Example
%----------------------------------------------------------------------------------------

\section{Software and Applied Example}
\label{software_section}

The present methods are implemented in an \texttt{R} package, \texttt{ConfoundedMeta}, which produces point estimates and inference for sensitivity analyses, tables across a user-specified grid of sensitivity parameters, and various plots. Descriptions of each function are provided in the Web Appendix and standard \texttt{R} documentation. A graphical user interface implementing the main functions is freely available (\url{https://mmathur.shinyapps.io/meta_gui_2/}). 

We illustrate the package's basic capabilities using an existing meta-analysis assessing, among several outcomes, the association of high versus low daily intake of soy protein with breast cancer risk among women \citep{trock}. The analysis comprised 20 observational studies that varied in their degree of adjustment for suspected confounders, such as age, body mass index (BMI), and other risk factors. To obtain $\tau^2_c$ and $\widehat{\text{Var}}(\tau^2_c)$ (which were not reported), we obtained study-level summary measures as reported in a table from \citet{trock}, approximating odds ratios with risk ratios given the rare outcome. This process is automated in the function \texttt{ConfoundedMeta::scrape\_meta}. We estimated $\widehat{y}_R^c = \log 0.82$, $\widehat{ \text{SE} } \left( \widehat{y}_R^c \right) = 8.8 \times 10^{-2}$ via the \citet{hartung-knapp} adjustment (whose advantages were demonstrated by \citet{inthout}), $\tau^2_c = 0.10$ via the \citet{paule} method, and $\widehat{ \text{SE} } \left( \tau^2_c \right) = 5.0 \times 10^{-2}$.

\begin{figure}[H]
\centering
\caption{Impact of varying degrees of unmeasured confounding bias on proportion of true relative risks $< 0.90$}
\label{fig:sensplot}
\includegraphics[width=100mm]{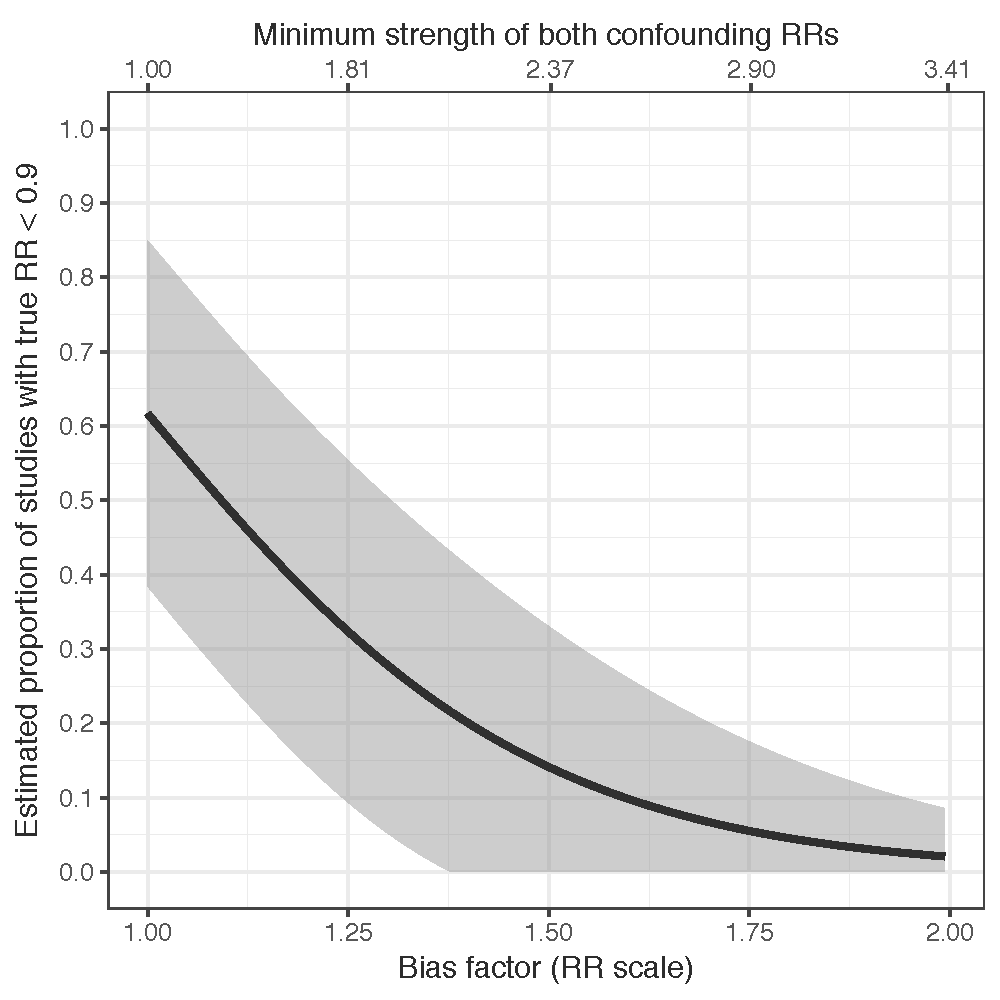}
\end{figure}

Figure \ref{fig:sensplot} (produced by \texttt{ConfoundedMeta::sens\_plot}) displays the estimated proportion of studies with true relative risks $< 0.90$ as a function of either the bias factor or the confounding strength, holding constant $\sigma^2_{B^{*}}=0.01$. Table \ref{T_table} (produced by \texttt{ConfoundedMeta::sens\_table}) displays $\widehat{T}(r,q)$ and $\widehat{G}(r,q)$ across a grid of values for $r$ and $q$. For example, only a bias factor exceeding $1.63$ on the relative risk scale (equivalently, confounding association strengths of $2.64$) could reduce to less than $10\%$ the proportion of studies with true relative risks $< 0.90$. However, variable bias across studies would reduce this proportion, and the confidence interval is wide.

%----------------------------------------------------------------------------------------
%	Simulation Study
%----------------------------------------------------------------------------------------
\section{Simulation Study}

We assessed finite-sample performance of inference on $\widehat{p}(q)$ in a simple simulation study. While fixing the mean and variance of the true effects to $\mu^t = \log 1.4$ and $V^t=0.15$ and the bias parameters to $\mu_{B^{*}} = \log 1.6$ and $\sigma^2_{B^{*}} = 0.01$, we varied the number of studies ($k \in \{ 15, 25, 50, 200 \}$) and the average sample size $N$ within each study ($E[N] \in \{ 300, 500, 1000\}$). The fixed parameters were chosen to minimize artifacts from discarding pathological samples with $\tau^2_c < \sigma^2_{B^{*}}$ or with truncated outcome probabilities due to extreme values of $RR_{XY}^c$. We ran $1000$ simulations for each possible combination of $k$ and $E[N]$, primarily assessing coverage of nominal 95\% confidence intervals and secondarily assessing their precision (total width) and bias in $\widehat{p}(q)$. 

For each study, we drew $N \sim \text{Unif}\left( 150, 2 E[N] - 150 \right)$, using $150$ as a minimum sample size to prevent model convergence failures, and drew the study's true effect size as $M^t \sim N(\mu^t, V^t)$. We simulated data for each subject under a model with a binary exposure $\left( X \sim \text{Bern}( 0.5 ) \right)$, a single binary unmeasured confounder, and a binary outcome. We set the two bias components equal to one another ($g = RR_{XU} = RR_{UY}$) and fixed $P(U=1 | X=1) = 1$, allowing closed-form computation of: $$P(U=1 | X=0) = \frac{ \exp(M^t) [ 1 + \left( g - 1 \right) ] - \exp(M^c) }{ (g-1) \exp(M^c) }$$
as in \citet{ding}. Within each stratum $X=x$, we simulated $U \sim \text{Bern}\left( P(U=1 | X=x) \right)$. We simulated outcomes as $Y \sim \text{Bern} \left( \exp \{ \log 0.05 + \log(g) U + M^t X \} \right)$. Finally, we computed effect sizes and fit the random-effects model using the \texttt{metafor} package in \texttt{R} \citep{viechtbauer}, estimating $\tau^2_c$ per \citet{paule} and $\widehat{\text{Var}}\left( \widehat{y}_R^c \right)$ with the \citet{hartung-knapp} adjustment.

Results (Table \ref{table:simres}) indicated approximately nominal performance for all combinations of $k$ and $E[N]$, with precision appearing to depend more strongly on $k$ than $E[N]$. As expected theoretically, $\widehat{p}(q)$ was approximately unbiased.

%----------------------------------------------------------------------------------------
%	Discussion
%----------------------------------------------------------------------------------------

\section{Discussion}

This paper has developed sensitivity analyses for unmeasured confounding in a random-effects meta-analysis of a relative risk outcome measure. Specifically, we have presented estimators for the proportion, $\widehat{p}(q)$, of studies with true effect sizes surpassing a threshold and for the minimum bias, $\widehat{T}(r,q)$, or confounding association strength, $\widehat{G}(r,q)$, in all studies that would be required to reduce to a threshold the proportion of studies with effect sizes less than $q$. Such analyses quantify the amount of confounding bias in terms of intuitively tractable sensitivity parameters. Computation of $\widehat{p}(q)$ uses two sensitivity parameters, namely the mean and variance across studies of a joint bias factor on the log-relative risk scale. Estimators $\widehat{T}(r,q)$ and $\widehat{G}(r,q)$ make reference to, and provide conclusions for, a single sensitivity parameter, chosen as either the common joint bias factor across studies or the strength of confounding associations on the relative risk scale. These methods assume that the bias factor is normally distributed or fixed across studies, but do not make further assumptions regarding the nature of unmeasured confounding.

Assessing sensitivity to unmeasured confounding is particularly important in meta-analyses of observational studies, where a central goal is to assess the current quality of evidence and to inform future research directions. If a well-designed meta-analysis yields a low value of $\widehat{T}(r,q)$ or $\widehat{G}(r,q)$ and thus is relatively sensitive to unmeasured confounding, this indicates that future research on the topic should prioritize randomized trials or designs and data collection that reduce unmeasured confounding. On the other hand, individual studies measuring moderate effect sizes with relatively wide confidence intervals may not, when considered individually, appear highly robust to unmeasured confounding; however, a meta-analysis aggregating their results may nevertheless suggest that a substantial proportion of the true effects are above a threshold of scientific importance even in the presence of some unmeasured confounding. Thus, conclusions of the meta-analysis may in fact be robust to moderate degrees of unmeasured confounding. 

We focused on relative risk outcomes because of their frequency in biomedical meta-analyses and their mathematical tractability, which allows closed-form solutions with the introduction of only one assumption (on the distribution of the bias factor). To allow application of the present methods, an odds ratio outcome can be approximated as a relative risk if the outcome is rare. If the outcome is not rare, the odds ratio can be approximately converted to a relative risk by taking its square root; provided that the outcome probabilities are between $0.2$ and $0.8$, this transformation is always within $25\%$ of the true relative risk \citep{tvw_odds_ratio}. Comparable sensitivity analyses for other types of outcomes, such as mean differences, would require study-level summary measures (for example, of within-group means and variances) and in some cases would yield closed-form solutions only at the price of more stringent assumptions. Under the assumption of an underlying binary outcome with high prevalence, such measures could be converted to log-odds ratios \citep{borensteinbook} and then to relative risks \citep{tvw_odds_ratio} as described above (see \citet{evalue}). It is important to note that, in circumstances discussed elsewhere \citep{tang,thorlund}, relative risk outcomes can produce biased meta-analytic estimates. When such biases in pooled point estimates or heterogeneity estimators are likely, sensitivity analyses will also be biased.

We operationalized ``robustness to unmeasured confounding'' as the proportion of true effects surpassing a threshold, an approach that focuses on the upper tail (for an apparently causative $RR_{XY}^c$) of the distribution of true effect sizes. Potentially, under substantial heterogeneity, a high proportion of true effect sizes could satisfy, for example, $RR_{XY}^t > 1.10$ while, simultaneously, a non-negligible proportion could be comparably strong in the opposite direction ($RR_{XY}^t < 0.90$). Such situations are intrinsic to the meta-analysis of heterogeneous effects, and in such settings, we recommend reporting the proportion of effect sizes below a symmetric threshold on the opposite side of the null (e.g., $\log 0.80$ if $q = \log 1.20$) both for the confounded distribution of effect sizes and for the distribution adjusted based on chosen bias parameters. For example, a meta-analysis that is potentially subject to unmeasured confounding and that estimates $\widehat{y}_R^c = \log 1.15$ and $\tau^2_c = 0.10$ would indicate that 45\% of the effects $RR_{XY}^c$ surpass $1.20$, while 13\% are less than $0.80$. For a common $B^{*} = \log 1.10$ (equivalently, $g = 1.43$), we find that $\left( 1 - \Phi \left( \frac{ \log 1.20 - \log 1.15 + \log 1.10 }{ \sqrt{0.10} } \right) \right) \cdot 100\% = 33\%$ of the true effects surpass $RR_{XY}^c = 1.20$, while 20\% are less than $RR_{XY}^c = 0.80$. More generally, random-effects meta-analyses could report the estimated proportion of effects above the null or above a specific threshold (along with a confidence interval for this proportion) as a continuous summary measure to supplement the standard pooled estimate and inference. Together, these reporting practices could facilitate overall assessment of evidence strength and robustness to unmeasured confounding under effect heterogeneity.

The proposed sensitivity analyses in theory require only standard summary measures from a meta-analysis (namely, the estimated pooled effect and a heterogeneity estimator to compute point estimates, along with their estimated variances to compute inference), rather than study-level data. However, in practice, we find that reporting of $\tau^2_c$ and $\widehat{ \text{Var} }\left( \tau^2_c \right)$ is sporadic in the biomedical literature. Besides their utility for conducting sensitivity analyses, we consider $\tau^2_c$ and $\widehat{ \text{Var} }\left( \tau^2_c \right)$ to be inherently valuable to the scientific interpretation of heterogeneous effects. We therefore recommend that they be reported routinely for random-effects meta-analyses, even when related measures, such as the proportion of total variance attributable to effect heterogeneity ($I^2$), are also reported. To enable sensitivity analyses of existing meta-analyses that do not report the needed summary measures, the package \texttt{ConfoundedMeta} helps automate the process of obtaining and drawing inferences from study-level data from a published forest plot or table. The user can then simply fit a random-effects model of choice to obtain the required summary measures.

Our framework assumes that the bias factor is normally distributed or taken to be fixed across studies. Normality is approximately justified if, for example, $RR_{XU}$ and $RR_{UY}$ are approximately identically and independently normal with relatively small variance. Since $RR_{UY}$ is in fact a maximum over strata of $X$ and the range of $U$, future work could potentially consider an extreme-value distribution for this component, but such a specification would appear to require a computational, rather than closed-form, approach. Perhaps a more useful, conservative approach to assessing sensitivity to bias that may be highly skewed is to report $\widehat{T}(r,q)$ and $\widehat{G}(r,q)$ for a wide range of fixed values $B^{*}$, including those much larger than a plausible mean.

An alternative sensitivity analysis approach would be to directly apply existing analytic bounds \citep{ding} to each individual study in order to compute the proportion of studies with effect sizes more extreme than $q$ given a particular bias factor. This has the downside of requiring access to study-level summary measures (rather than pooled estimates). Moreover, the confidence interval of each study may be relatively wide, such that no individual study appears robust to unmeasured confounding, while nevertheless a meta-analytic estimate that takes into account the distribution of effects may in fact indicate that some of these effects are likely robust. One could also alternatively conduct sensitivity analyses on the pooled point estimate itself, but such an approach is na{\"i}ve to heterogeneity: when the true effects are highly variable, a non-negligible proportion of large true effects may remain even with the introduction of enough bias to attenuate the pooled estimate to a scientifically unimportant level.

In summary, our results have shown that sensitivity analyses for unmeasured confounding in meta-analyses can be conducted easily by extending results for individual studies. These methods are straightforward to implement through either our \texttt{R} package \texttt{ConfoundedMeta} or graphical user interface and ultimately help inform principled causal conclusions from meta-analyses.

%----------------------------------------------------------------------------------------
%	RESEARCH TRANSPARENCY
%----------------------------------------------------------------------------------------
\section*{Reproducibility}
All code required to reproduce the applied example and simulation study is publicly available (\url{https://osf.io/2r3gm/}).

%----------------------------------------------------------------------------------------
%	SUPPLEMENTARY
%----------------------------------------------------------------------------------------
\section*{Supplementary Materials}
Web Appendices referenced in Sections \ref{main_results_section} and \ref{software_section} are available with this paper at the \emph{Biometrics} website on Wiley Online Library.

%----------------------------------------------------------------------------------------
%	ACKNOWLEDGMENTS
%----------------------------------------------------------------------------------------
\section*{Acknowledgments}
This research was supported by National Defense Science and Engineering Graduate Fellowship 32 CFR 168a and NIH grant ES017876.

%----------------------------------------------------------------------------------------
%	REFERENCE LIST
%----------------------------------------------------------------------------------------
%\pagebreak
\bibliography{refs.bib}
\bibliographystyle{apacite}
%\bibliographystyle{plain}

%----------------------------------------------------------------------------------------
%	Tables and Figures
%----------------------------------------------------------------------------------------
\newpage
\section*{Tables and Figures}

\begin{table}[H]
\centering
\caption{Bounds on $\widehat{p}(q)$ provided by homogeneous bias with an apparently causative or preventive pooled effect. $\widehat{\mu}^t$ estimates $\mu^t$ and is equal to $\widehat{y}_R^c - \mu_{B^{*}}$ for $\widehat{y}_R^c > 0$ or $\widehat{y}_R^c + \mu_{B^{*}}$ for $\widehat{y}_R^c < 0$.}
\label{homo-bias}
\begin{tabular}{@{}lllll@{}}
\toprule
\centering
           & $q > \widehat{\mu}^t$ & $q < \widehat{\mu}^t$   &  &  \\ \midrule
$\widehat{y}_R^c > 0$  & Upper bound  & Lower bound &  &  \\
$\widehat{y}_R^c < 0$ & Lower bound  & Upper bound &  &  \\ \bottomrule
\end{tabular}
\end{table}

\clearpage

\clearpage
\begin{table}[H]
\centering
\caption{$\widehat{T}(r,q)$ and $\widehat{G}(r,q)$ (in parentheses) for varying $r$ and $q$. Blank cells indicate combinations for which no bias would be required.}
\label{T_table}
  \begin{tabular}{lSSSSSS}
    \toprule
    \multirow{3}{*}{$r$} &
      \multicolumn{3}{c}{$q$} & \\
      & {0.70} & {0.80} & {0.90} \\
      \midrule
	0.1 & {1.27 (1.85)} & {1.45 (2.25)} & {1.63 (2.64)} \\ 
  0.2 & {1.10 (1.44)}  & {1.26 (1.84)} & {1.42 (2.19)}  \\ 
  0.3 &  & {1.14 (1.55)} & {1.29 (1.89)} \\ 
  0.4 &  & {1.05 (1.28)} & {1.18 (1.64)} \\ 
  0.5 &  &  & {1.09 (1.41)} \\ 
    \bottomrule
  \end{tabular}
\end{table}

\clearpage
\begin{table}[H]
\centering
\caption{Point estimate bias, 95\% confidence interval (CI) coverage, and 95\% CI width for varying numbers of studies ($k$) and mean sample sizes within each study (Mean $N$).}
\label{table:simres}
\begin{tabular}{rrrrr}
  \toprule
$k$ & Mean $N$ & $\widehat{p}$ bias & CI coverage & CI width \\ 
  \midrule
15 & 300 & 0.030 & 0.968 & 0.572 \\ 
  25 & 300 & 0.034 & 0.976 & 0.452 \\ 
  50 & 300 & 0.031 & 0.967 & 0.315 \\ 
  200 & 300 & 0.028 & 0.929 & 0.154 \\ \midrule
  15 & 500 & 0.022 & 0.967 & 0.524 \\ 
  25 & 500 & 0.022 & 0.977 & 0.408 \\ 
  50 & 500 & 0.025 & 0.974 & 0.283 \\ 
  200 & 500 & 0.024 & 0.934 & 0.140 \\ \midrule
  15 & 1000 & 0.018 & 0.976 & 0.479 \\ 
  25 & 1000 & 0.016 & 0.976 & 0.370 \\ 
  50 & 1000 & 0.018 & 0.969 & 0.259 \\ 
  200 & 1000 & 0.015 & 0.970 & 0.129 \\ 
   \bottomrule
\end{tabular}
\end{table}

%----------------------------------------------------------------------------------------

%\end{multicols}

\end{document}